\documentclass[prl,twocolumn,superscriptaddress,fleqn]{revtex4-2}

\usepackage{amsmath,amssymb}
\usepackage{graphicx}
\usepackage{epstopdf}
\usepackage{bm}
\usepackage{xcolor}
\newcommand{\NPS}     {NiPS$_3$}

\newcommand{\pho}	{$^{31}$P}
\newcommand{\sul}	{$^{33}$S}
\newcommand{\slr} 	{$T_1^{-1}$}
\newcommand{\slrP} 	{$^{31}T_1^{-1}$}
\newcommand{\slrS} 	{$^{33}T_1^{-1}$}

\newcommand{\hf} 	{$A_\text{hf}$}

\newcommand{\kk} 	{$\mathcal{K}$}

\begin{document}

\title[]{Microscopic evidence for a Zhang-Rice triplet state in the van der Waals antiferromagnet, NiPS$_3$}

\author{Beom Hyun Kim}
\thanks{These authors contributed equally to this work.}
%\email[]{bomisu@gmail.com}
%\homepage[]{Your web page}
\affiliation{Center for Quantum Materials, Seoul National University, Seoul 08826, Republic of Korea}
\affiliation{Department of Physics \& Astronomy, Seoul National University, Seoul 08826, Republic of Korea}
\author{Youjin Lee} 
\thanks{These authors contributed equally to this work.}
%\email[]{lyj3199@snu.ac.kr}
%\homepage[]{Your web page}
\affiliation{Center for Quantum Materials, Seoul National University, Seoul 08826, Republic of Korea}
\affiliation{Department of Physics \& Astronomy, Seoul National University, Seoul 08826, Republic of Korea}
\author{Junik Hwang}
%\email[]{wnsdlr0217@gmail.com}
%\homepage[]{Your web page}
\affiliation{Department of Physics, Changwon National University, Changwon 51139, Republic of Korea}
%\author{Seonghoon Park}
%\email[]{digh10005@gmail.com}
%\homepage[]{Your web page}
%\affiliation{Department of Physics, Changwon National University, Changwon 51139, Republic of Korea}
\author{Junghyun Kim}
%\email[]{jhyun0915@snu.ac.kr}
%\homepage[]{Your web page}
\affiliation{Center for Quantum Materials, Seoul National University, Seoul 08826, Republic of Korea}
\affiliation{Department of Physics \& Astronomy, Seoul National University, Seoul 08826, Republic of Korea}
\author{Je-Geun Park}
\email[]{jgpark10@snu.ac.kr}
%\homepage[]{Your web page}
\affiliation{Center for Quantum Materials, Seoul National University, Seoul 08826, Republic of Korea}
\affiliation{Department of Physics \& Astronomy, Seoul National University, Seoul 08826, Republic of Korea}
\affiliation{Institute of Applied Physics, Seoul National University, Seoul 08826, Republic of Korea}
\author{Seung-Ho Baek}
\email[]{sbaek.fu@gmail.com}
%\homepage[]{Your web page}
\affiliation{Department of Physics, Changwon National University, Changwon 51139, Republic of Korea}
%\affiliation{Department of Materials Convergence and System Engineering, Changwon National University, Changwon 51139, Korea}

\date{\today}

%%%%%%%%%%%%%%%%%%%%%%%%%%%%%%%%%%%%%%%%%%%%%%%%%%%%%%%%%%%%%%%%%%%%

\begin{abstract}

Quantum-entangled states underpin many emergent phenomena in quantum materials, yet their direct experimental identification remains a challenge. \NPS, a van der Waals antiferromagnet exhibiting a resolution-limited magnetic exciton in its ordered phase, has been proposed to host a many-body entangled Zhang-Rice triplet state. Here, using \sul\ nuclear magnetic resonance (NMR) on \sul-enriched \NPS\ single crystals, we provide microscopic evidence for this charge-transfer state. The \sul\ and \pho\ Knight shifts as a function of temperature reveal a unified spin-triplet configuration arising from strong hybridization between a self-doped hole in the S $3p$ orbitals and a hole in Ni $3d$ orbitals. Furthermore, the \sul\ nuclear spin-lattice relaxation rate exhibits a power-law divergence as it approaches the N\'eel temperature $T_N=155$ K, indicating critical slowing down of collective charge fluctuations consistent with spin-nematic correlations. These results reveal a spin-charge-intertwined ground state and establish the microscopic foundation for the exceptional coherence of the magnetic exciton in \NPS. 

\end{abstract}

%\keywords{Nuclear magnetic resonance, charge nematic, first-order phase transition, van der Waals magnets}

\maketitle

%%%%%%%%%%%%%%%%%%%%%%%%%%%%%%%%%%%%%%%%%%%%%%%%%%%%%%%%%%%%%%%%%%%%
Entanglement lies at the heart of quantum mechanics. Yet, its understanding and experimental verification remain profoundly challenging, as exemplified by the long-standing debate surrounding the Einstein-Podolsky-Rosen paradox \cite{reid09}. Even in the current era of quantum science and technology, the experimental identification of new quantum-entangled states in solids is exceedingly rare. Conceptually, the simplest entangled states are spin-singlet and spin-triplet configurations, first discussed in the Heitler-London theory of the hydrogen molecule \cite{ashcroft}. In correlated electron systems, a modern realization of such entanglement is the Zhang-Rice state, originally proposed to describe the emergent many-body physics of high-temperature copper-oxide superconductors \cite{zhang88}.

\NPS\ is among the most extensively studied van der Waals antiferromagnets \cite{kang20,hwangbo21,wang21,dirnberger22,klaproth23,he24,hamad24,sun24,jana25}. It has a monoclinic crystal structure with a space group $C2/m$ [Fig.\,1(a)] with the Ni$^{2+}$ ions arranged in a honeycomb lattice \cite{ouvrard85}.
It undergoes a zigzag magnetic ordering below the N\'eel temperature $T_N=155$ K \cite{kim18a,wildes22}, which breaks the three-fold rotational symmetry. Remarkably, \NPS\ exhibits an excitonic mode in its magnetically ordered phase that sharpens dramatically upon cooling and becomes resolution-limited at low temperatures, achieving an unprecedented linewidth of approximately 0.4 meV below 50 K \cite{kang20}. This unusual temperature evolution indicates that the spin degree of freedom plays a crucial role in the formation of the exciton, motivating its designation as a magnetic exciton. Consistent with this interpretation, the exciton shows a pronounced thickness dependence: it broadens significantly when the thickness is reduced to below three layers and disappears entirely in the monolayer limit \cite{kang20}, underscoring intimate coupling between magnetism and the exciton.

Since the discovery of the magnetic exciton in  \NPS, a variety of theoretical scenarios have been proposed, including a Zhang-Rice exciton \cite{kang20}, a triplet-singlet exciton \cite{klaproth23}, a Hund's exciton \cite{he24}, a polaron exciton \cite{hamad24}, and a 
spin-entangled exciton \cite{jana23, jana25}. While each of these models appears to capture certain aspects of the experimental observations, a coherent, unified microscopic picture has remained elusive. In particular, definitive experimental evidence identifying the nature of the underlying quantum-entangled ground state has been lacking, leaving the origin of the exceptional exciton coherence unresolved.

In this work, we present a \sul\ nuclear magnetic resonance (NMR) study of \sul-enriched \NPS\ single crystals in the paramagnetic (PM) state. Detailed analysis of temperature- and angle-dependent, quadrupole-perturbed \sul\ spectra reveals substantial transferred spin density on the S $3p$ orbitals and strong hybridization with Ni $3d$ moments---both hallmarks of a self-doped Zhang-Rice triplet ground state. A striking observation is the power-law divergence of the \sul\ nuclear spin-lattice relaxation rate \slr\ approaching $T_N$. Remarkably, this divergent behavior is absent in the \pho\ relaxation data. By comparing the two nuclei, we conclude that the fluctuations probed  by \sul\ originate from nematic correlations of the Ni moments. These results provide a clear microscopic perspective on the intertwined spin and charge degrees of freedom in this van der Waals antiferromagnet.

Single crystals of NiPS$_3$ enriched with \sul\ were synthesized using the chemical vapor transport (CVT) method. Stoichiometric amounts of Ni powders (Alfa Aesar, 99.99\%), P powder (Sigma-Aldrich, 99.99\%) and \sul\ (Sigma-Aldrich, 99\%) were mixed, with an additional 10 wt \% excess of \sul. All starting materials were handled and mixed in an Ar-filled glovebox to prevent oxidation. The mixture was sealed in an evacuated quartz ampoule and placed in a two-zone furnace for 7 days, the hot and cold zones maintained at 750 $^\circ$C and 720 $^\circ$C, respectively. The chemical composition of the grown crystals was confirmed by energy-dispersive X-ray spectroscopy using a COXEM-EM30 scanning electron microscope equipped with a Bruker QUANTAX 70
energy-dispersive X-ray system. The crystal structure was confirmed through XRD measurements using a commercial diffractometer (Rigaku Miniflex II). Additionally, we verified that the bulk magnetic susceptibility of the \sul-enriched crystal is identical to that of the unenriched one.

$^{33}$S (nuclear spin $I=3/2$) NMR was carried out in a \sul\ enriched \NPS\ single 
crystal ($3\times2\times0.2$ mm$^3$) at an external field of 5.93 T and in 
the range of temperature 150 --- 320 K. %For Cd-doped \NPS\ crystals, \pho\ NMR was carried out in the same external field. 
The sample was rotated using a goniometer for the exact alignment along the external field. The \sul\ NMR  spectra were acquired by a standard spin-echo technique with a typical $\pi/2$ pulse length 2--3 $\mu$s. 
The \sul\ nuclear spin-lattice relaxation rate \slr\ was obtained by fitting the 
recovery of the nuclear magnetization $M(t)$ after a saturating pulse to the following fitting function, $1-M(t)/M(\infty)=A [0.1\exp(-t/T_1)+0.9\exp(-6t/T_1)]$ where $A$ is a fitting parameter that is ideally unity. NMR measurements were performed exclusively in the PM state, because the signal is lost below $T_N$ due to substantial magnetic broadening.

\begin{figure}
\centering
\includegraphics[width=\linewidth]{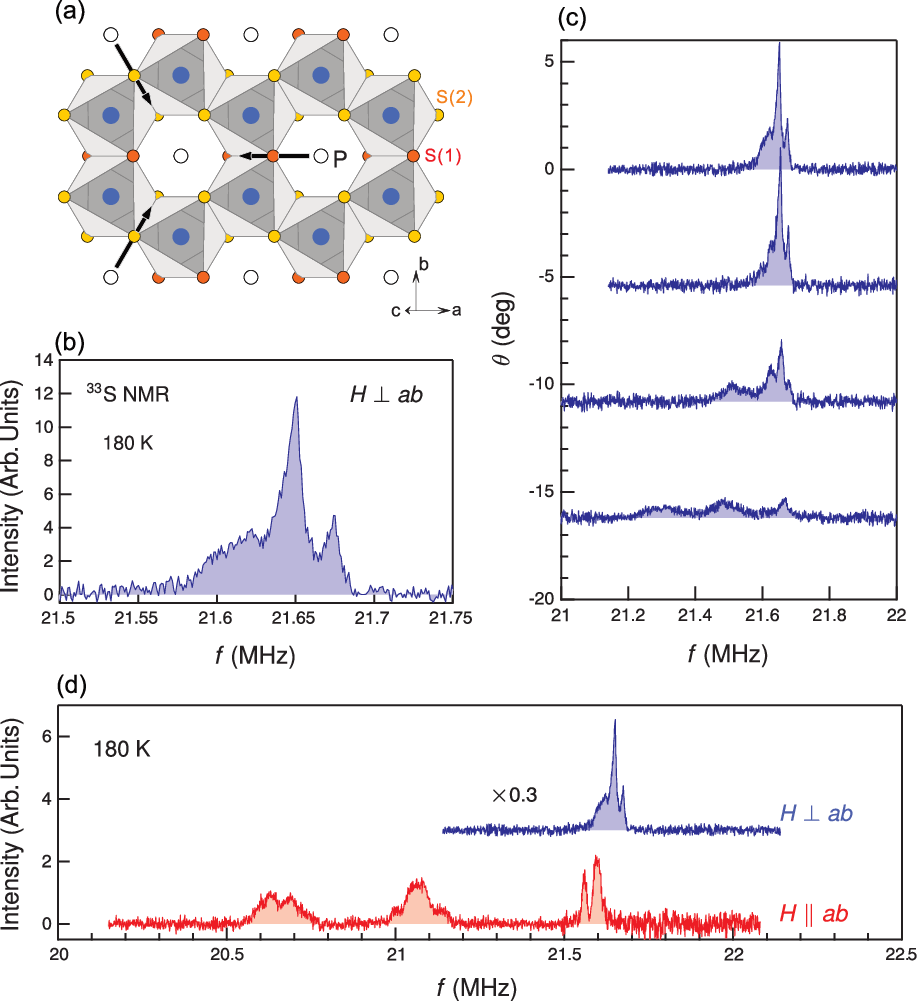}
\caption{(a) Crystal structure of \NPS\ viewed along the $c^*$ ($\perp ab$) axis. The black arrows denote the three different directions of the principal components $V_{zz}$ of the local EFG tensor at sulfur sites. (b) Central transition of the \sul\ NMR spectrum measured at 180 K in an external field of $H=5.93$ T applied perpendicular to the $ab$ plane. (c) Angular dependence of the $^{33}$S spectrum with respect to $c^*$ ($H \perp ab$). The spectral lines spread and broaden rapidly with increasing $\theta$.
(d) $^{33}$S NMR spectra for $H\parallel ab$ ($\theta=90^\circ$). The three distinct groups of lines are associated with the three different $V_{zz}$ directions with respect to $\mathbf{H}$. 
}\label{S_angle}
\end{figure}

The \sul\ NMR central transition ($1/2 \leftrightarrow -1/2$) was found in \sul -enriched \NPS\ single crystals at an external field $H= 5.93$ T applied perpendicular to the $ab$ plane ($H\perp ab$ or $H\parallel c^*$). Figure \ref{S_angle}(b) shows a representative \sul\ spectrum measured at 180 K, consisting of two sharp lines and one broad feature. 
As the angle $\theta$ between the $c^*$ axis and $\mathbf{H}$ increases, the spectrum broadens rapidly and evolves into a complicated structure, as shown in Fig.\,\ref{S_angle}(c). For $\theta \gtrsim 15^\circ$, tracing the angular evolution of individual lines becomes exceedingly difficult. We therefore aligned the $ab$ plane parallel to $\mathbf{H}$ ($\theta=90^\circ$) and obtained three distinct groups of \sul\ resonance lines, each containing 2--4 components [Fig.\,\ref{S_angle}(d)]. 
Because \sul\ has nuclear spin $I=3/2$ and sizable quadrupole moment $Q$, the central transition experiences a second-order quadrupole shift $\Delta\nu_\text{2nd}$ in the presence of an electric field gradient (EFG). Based on local site symmetry, the principal component $V_{zz}$ of the local EFG tensor at the S nuclei is expected to point perpendicular to the NiS$_6$ octahedral edges, tilting out of the $ab$ plane by $\sim 30^\circ$. 
Indeed, the three-fold lattice symmetry and associated quadrupole interactions qualitatively account for the three distinct resonance groups observed for $H\parallel ab$, since the angle between $V_{zz}$ and the external field differs among the three \sul\ sites. %In addition, a possible in-plane anisotropy of the Knight shift may further contribute to the observed line distribution  for $H\parallel ab$.
For $H\parallel c^*$, by contrast, the local symmetries of the three \sul\ sites yield the identical resonant spectrum, thereby explaining the nearly collapsed spectral lines. The residual line splitting may arise from a slight sample misalignment and/or small differences in EFG magnitude between the S(1) and S(2) sites. 
A detailed analysis of the angle dependence of the \sul\ lines reveals that the $V_{zz}$ components actually lie within the $ab$ plane, as denoted by the black arrows in Fig.\,\ref{S_angle}(a). 
As detailed in the Supplemental Note 1 and Fig.\,S1 \cite{supple1}, the angle dependence of the \sul\ spectrum is well reproduced by a quadrupole frequency $\nu_Q= 6.4$ MHz and an anisotropy parameter $\eta=0.1$, with the three-fold $V_{zz}$ orientations confined within the $ab$ plane. 
The extracted $\nu_Q=6.4$ MHz substantially exceeds values typical of layered transition-metal disulfides \cite{sutrisno09} and is comparable to that of the molecular ${}^2\Pi$ SH system \cite{miller71}. This unusually large quadrupole interaction is hardly understood within a purely localized $d^8$ description of the Ni$^{2+}$ ions, suggesting strong hybridization of the S $3p$ holes with asymmetrically arranged surrounding Ni $3d$ orbitals.
 
\begin{figure*}
\centering
\includegraphics[width=0.65\linewidth]{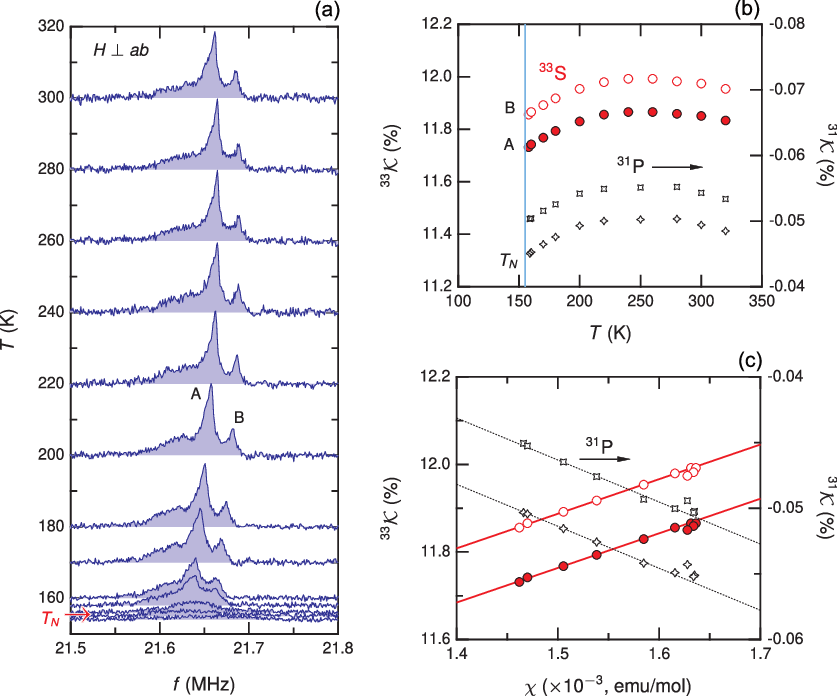}
\caption{(a) Temperature dependence of the $^{33}$S NMR spectrum in the PM phase for $H \perp ab$. The overall spectral shape remains unchanged across the temperature range, with two well-resolved peaks (labelled A and B) persisting up to room temperature.  (b) \sul\ Knight shift $^{33}\mathcal{K}$ of the A and B peaks as a function of temperature. The temperature dependence of both $^{33}\mathcal{K}$ and \pho\ Knight shift $^{31}\mathcal{K}$ (right axis) closely follows that of the bulk spin susceptibility, indicating that both nuclei probe the identical local spin susceptibility. Note that the two lines for \pho\ arise from nuclear dipole-dipole interactions, so called Pake doublet \cite{pake48a}.
(c) Knight shift versus susceptibility (\kk -$\chi$) plots with temperature as an implicit parameter yield hyperfine coupling constants of $^{33}A_\text{hf}=44.2$ kOe/$\mu_B$ for \sul\ and $^{31}A_\text{hf}=-1.8$ kOe/$\mu_B$ for \pho. % evidencing the considerable spin density at S $3p$ transferred from Ni $3d$. %The temperature-independent orbital contributions to the Knight shifts are $^{33}\mathcal{K}_\text{orb} = 10.7$ \% and $^{31}\mathcal{K}_\text{orb} = -0.004$ \%, respectively.
}
\label{S_spec}
\end{figure*} 

We measured the temperature dependence of the \sul\ spectrum for $H\parallel c^*$ [Fig.\,\ref{S_spec}(a)]. Apart from the rapid suppression of signal intensity near $T_N$, attributable to the \textit{wipeout effect} caused by the shortening of NMR relaxation times, the overall spectral shape remains essentially unchanged up to room temperature. In particular, the sharp peaks labelled A and B persist over the whole temperature range, enabling reliable determination of the \sul\ Knight shift, $^{33}\mathcal{K}$  [Fig.\,\ref{S_spec}(b)]. Here, $\mathcal{K}$ is defined as $(\nu-\nu_0)/\nu_0\times 100\%$ where $\nu_0$ is the unshifted Larmor frequency and $\nu$ is the resonance frequency corrected for the second-order quadrupole shift.
The temperature dependence of $^{33}\mathcal{K}$ resembles that of the \pho\ Knight shift $^{31}\mathcal{K}$ \cite{hwang25}, with both exhibiting a broad maximum near 250 K, which is consistent with bulk susceptibility \cite{joy92,wildes15}. This demonstrates that both \sul\ and \pho\ nuclei probe the same local spin susceptibility associated with the Ni moments.

The spin part of the Knight shift is proportional to the local spin susceptibility, $\mathcal{K}_\text{spin} = A_\text{hf}\chi_\text{spin}$. The hyperfine coupling constant \hf\ can be extracted from the \kk -$\chi$ plot with temperature as an implicit parameter \cite{clogston64a}, as shown in Fig.\,\ref{S_spec}(c). A linear fit below 250 K yields $^{33}{A}_\text{hf}=44.2 \text{ kOe}/\mu_B$. For \pho, the corresponding value is $^{31}A_\text{hf}=-1.8$ kOe/$\mu_B$. % and $^{31}\mathcal{K}_\text{orb} = -0.004$ \%.  
The breakdown of linearity above 250 K for both nuclei further supports their common origin in the Ni spin susceptibility. 
Since the calculated $^{33}{A}_\text{hf}$ from Ni moments  is small and comparable to $^{31}{A}_\text{hf}$ (see Supplemental Note 2 \cite{supple1}), the much larger hyperfine coupling at \sul\ than that at \pho\ implies unusually high spin density on the otherwise completely filled S $3p$ orbitals. This shows that $^{33}\mathcal{K}$ primarily probes S $3p$ moments (holes) transferred from Ni $3d$ orbitals (see Supplemental Note 3 \cite{supple1}).
Since $^{31}\mathcal{K}$ reflects the Ni $3d$ moments via dipolar coupling, the identical temperature dependence of both $^{31}\mathcal{K}$ and $^{33}\mathcal{K}$ demonstrates that the S $3p$ and Ni $3d$ moments are strongly coupled, acting as components of a unified spin system. These results suggest that the spin-triplet $\mathrm{^3A_2}$ symmetry of Ni$^{2+}$ ($d^8$) ions is preserved in the charge-transfer state with $d^9\underbar{L}$ configuration, where $\underbar{L}$ denotes a ligand hole.
The ground state of \NPS\ is therefore best described as a $p$-$d$ hybridized spin-triplet state, characterized by the superposition of $d^8$ and $d^9\underbar{L}$ configurations, namely a Zhang-Rice triplet (ZRT) state \cite{kang20,kim23a,song24}, rather than a state with dominant $d^8$ character. 
Even within the Hund's exciton scenario \cite{he24}, which primarily focuses on a $d^8$-based description, a substantial $d^9\underbar{L}$ ligand-hole admixture of approximately 40\% was estimated.
Note that, since the ZRT represents the local electronic configuration of \NPS, it remains a valid description in both the PM and AFM phases, with the latter corresponding to the long-range alignment of these local triplet many-body states.
Consequently, all alternative exciton scenarios \cite{klaproth23, hamad24, jana25}, which are essentially based on the localized $d^8$ configuration, are incompatible with these NMR results. 

An unexpected finding in Fig.\,\ref{S_spec}(c) is the unprecedentedly large residual NMR shift of 10.7 \%. This shift includes the second-order quadrupole contribution, approximately 2 \% for $\nu_Q=6.4$ MHz, implying that the additional 8.7 \% shift ought to be attributed to the orbital Knight shift, $\mathcal{K}_\text{orb}$. In general, $\mathcal{K}_\text{orb}$ is related to the Van Vleck susceptibility $\chi_\text{VV}$ via $\mathcal{K}_\text{orb}=2\langle r^{-3}\rangle /N_A \chi_\text{VV}$ \cite{clogston64a} where $N_A$ is the Avogadro's number and  $\langle r^{-3}\rangle$ is the expectation value of $r^{-3}$. Using $\langle r^{-3}\rangle_{3p}=4.8$ a.u.\,for a $3p$ orbital \cite{morton64}, we obtain $\chi_\text{VV}\approx 0.8\times 10^{-3}$ emu/mol.
In an octahedral crystal field, Ni$^{2+}$ ions possess a $\mathrm{^3A_2}$ orbital-singlet ground state and a $\mathrm{^3T_2}$ orbital-triplet first-excited state. Their mixing can produce a finite orbital susceptibility depending on the excitation gap $\Delta$.
%(Ref.\,\cite{griffith57}): 
%\begin{equation}
%	\chi_\text{VV} = 16\frac{N_A\mu_B^2}{\Delta}.
%\end{equation}
However, the estimated $\chi_\text{VV}$ requires the excitation gap $\Delta$ to be an order of magnitude smaller than the calculated value of 1 eV \cite{kang20} (see Supplemental Note 4 \cite{supple1}). %Thus, the origin of the large orbital Knight shift remains unresolved. While intriguing, this issue does not affect our main conclusions and is left as an open question for future study. 
The large orbital Knight shift thus suggests contributions beyond the standard local crystal-field model, possibly arising from the significant inter-site orbital currents inherent to the hybridized ZRT state. 

\begin{figure}
\centering
\includegraphics[width=0.85\linewidth]{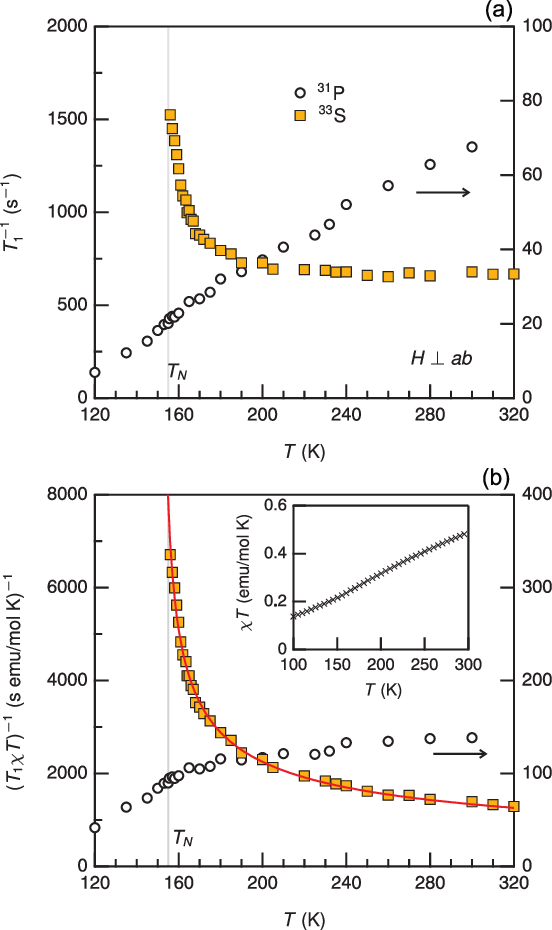}
\caption{(a) Nuclear spin-lattice relaxation rate \slr\ of \sul\ and \pho\ as a function of temperature. Upon cooling, while $^{31}T_1^{-1}$ monotonically decreases, $^{33}T_1^{-1}$ exhibits a progressive enhancement. 
(b) Temperature dependence of the reduced relaxation rate, $R_1\equiv (T_1\chi T)^{-1}$. While the nearly constant $^{31}R_1$ supports the absence of collective spin fluctuations, $^{33}R_1$ exhibits a power-law divergence, which is well-fitted by Eq.\,(1) throughout the entire temperature range (solid line). This divergence demonstrates the development of collective charge fluctuations towards an incipient nematic ordering instability (see text). Inset : Temperature dependence of $\chi T$ for $H\perp ab$. 
}
\label{S_T1}
\end{figure}

Having established the ZRT ground state in the PM phase of \NPS, we now turn to the low-energy spin/charge dynamics probed by the \sul\ nuclear spin-lattice relaxation rate, \slrS\ [Fig.\,\ref{S_T1}(a)]. Measurements were performed at 5.93 T applied along the $c^*$ axis and compared with the \pho\ data obtained under the same field orientation \cite{hwang25}. 
Upon cooling, \slrS\ increases, exhibiting a divergent behavior as $T_N$ is approached, in sharp contrast to the monotonic decrease observed for \slrP, which indicates the absence of collective spin fluctuations above $T_N$. 
The contrasting relaxation behavior may result from geometric filtering \cite{shastry89}, if the hyperfine form factor $A(\mathbf{q})$ vanishes at the P site. However, our calculations (see Supplemental Note 2 \cite{supple1}) show that $A(\mathbf{q})$ remains finite for all relevant $\mathbf{q}$, ruling out a symmetry-imposed cancellation of AFM fluctuations. 
%Since both nuclei probe the same spin susceptibility [Fig.\,\ref{S_spec}(b)], 
Consequently, the divergence in \slrS\  must stem from fluctuations of the EFG, which couple to quadrupolar \sul\ ($I=3/2$), but not to \pho\ ($I=1/2$). In \NPS, such EFG fluctuations are naturally linked to charge dynamics involving Ni $3d$ and S $3p$ charge transfer.

We note that the suppression of \slrP\ with decreasing temperature follows the temperature dependence of $\chi T$ [see the inset of Fig.\,\ref{S_T1}(b)], suggesting that static short-range AFM correlations dominate the spin contribution to \slr\ in the PM phase. 
To isolate the contribution of dynamic spin correlations, we define the reduced spin-lattice relaxation rate \cite{baek04} $R_1 \equiv (T_1 \chi T)^{-1}$. This correction yields an almost temperature-independent $^{31}R_1$ down to $T_N$, confirming the absence of critical spin dynamics. 
In striking contrast, $^{33}R_1$ increases rapidly on cooling and follows a power-law divergence, 
\begin{equation}
	^{33}R_1\propto |T-T_0|^{-0.464}.
\end{equation}
The fitting of the data yields the characterization temperature $T_0=152$ K, which is remarkably close to $T_N=155$ K [solid line in Fig.\,\ref{S_T1}(b)]. 
%We stress that, although the characteristic temperature $T_0$ is only slightly below $T_N=155$ K, the divergence of $^{33}R_1$ is abruptly truncated at $T_N$.  The power-law growth of $^{33}R_1$ therefore emerges from an incipient instability, rather than the AFM transition itself. 
 %
 
% Discussion ---

We now like to put our NMR findings into a unified picture. The \sul\ and \pho\ Knight shift measurements demonstrate substantial transferred spin density on S $3p$ orbitals and its strong coupling with Ni $3d$ moments, consistent with the formation of a self-doped ZRT ground state in the PM phase. The pronounced $p$-$d$ hybridization intrinsic to the ZRT state naturally explains the unusually large quadrupole frequency $\nu_Q=6.4$ MHz observed in our measurements. 
Surprisingly, despite the dominant spin character of the ZRT state, the low-energy spin dynamics is effectively inactive in the nuclear spin-lattice relaxation rate measurements. Instead, the \sul\ relaxation rate exhibits a power-law divergence toward $T_N$, suggesting a long-range ordering instability that predominantly involves charge fluctuations, which is distinct from the zig-zag AFM ordering. 
%
%Because the self-doped ligand-hole concentration is extremely low, the emergence of a conventional charge-density wave is unlikely. 

The contrasting behaviors of spin and charge fluctuation spectra may originate from quantum fluctuations enhanced by biquadratic spin interactions \cite{scheie23, mellado24}. 
Given the strong spin-charge coupling inherent to the ZRT state \cite{kim18a,kang20}, the hole density on a given S site bridging two Ni sites is dependent on the relative spin arrangement. Namely, it is slightly higher for ferromagnetic than for antiferromagnetic coupling \cite{zhang25}, due to the Hund's coupling on the ligand orbitals. The corresponding coupling can be expressed as $\bar{n}_{ij} = \bar{n}_0 + \bar{n}_1 \left(\textbf{S}_i \cdot \textbf{S}_j \right)$, where $\bar{n}_0$ and $\bar{n}_1$ denote the spin-independent and spin-dependent contributions of the self-doped S holes, respectively.
The correlation function of charge fluctuations, $C_{\bar{n}_{ij},\bar{n}_{kl}} = \left<\bar{n}_{ij} \bar{n}_{kl} \right> - \left<\bar{n}_{ij} \right> \left<\bar{n}_{kl} \right>$, is therefore governed by the nematic correlations of Ni moments, since $C_{\bar{n}_{ij},\bar{n}_{kl}}=\bar{n}_1^2 \left[ \left< \left( \textbf{S}_i \cdot \textbf{S}_j \right) \left( \textbf{S}_k \cdot \textbf{S}_l \right) \right> -  \left< \textbf{S}_i \cdot \textbf{S}_j \right> \left< \textbf{S}_k \cdot \textbf{S}_l \right>  \right]$. 
Consequently, the observed divergence of $^{33}R_1$ can be interpreted as a manifestation of critical nematic fluctuations that couple to the charge degrees of freedom through spin-orbit entanglement.
Although no experimental signature of nematic order has yet been observed in bulk material, a vestigial nematic phase has been reported in the single-layer limit, where strong quantum fluctuations suppress the long-range AFM order \cite{tan24,sun24}. By analogy, in the bulk system, thermal fluctuations in the quasi-two-dimensional lattice may destroy long-range spin and nematic order while allowing a nematic instability to persist within individual layers. In this scenario, interlayer decoupling prevents the establishment of true long-range order, but robust intralayer nematic correlations survive above $T_N$. While bulk probes like neutron scattering barely detect 2D fluctuations, NMR, as a local probe in real space, is uniquely sensitive to such fluctuating 2D nematicity.

However, this nematic instability is preempted by the magnetic transition at $T_N$. 
We propose that once the interlayer nematic correlations become sufficiently strong, the enhanced spin-charge coupling drives the spin alignment along the nematic orientation, thereby stabilizing the zigzag AFM order at $T_N$.  
Such a nematicity-driven magnetic ordering scenario naturally accounts for both the first-order AFM transition revealed by our \pho\ NMR study \cite{hwang25} and the unusual tilt of the ordered moments toward S(1) \cite{wildes15}. 
The recent observation of nematicity within the AFM phase of isostructural FePSe$_3$ under uniaxial strain \cite{hwangbo24,yao26} further supports the intimate link between nematic and magnetic orders, suggesting that this interplay  may be a general feature of van der Waals antiferromagnets. 

\begin{acknowledgements}
	The work at SNU was supported by the Samsung Science \& Technology Foundation (Grant No.~SSTFBA2101-05) and the Leading Researcher Program of the National Research Foundation of Korea (Grant
No.~RS-2020-NR049405).
\end{acknowledgements}

\paragraph{Data Availability.}The data that supports the findings of this article are available upon reasonable request from the authors.

\bibliography{mybib}

%\item[Author Contributions] JGP and SHB have proposed and initiated the
%	project.  YL and JK have grown \sul\ enriched \NPS\ single crystals and carried out all characterization, and BHK carried out theoretical calculations. JH and SHB performed NMR
%		measurements and analyzed data. SHB, JGP, and BHK participated in writing of the manuscript. All authors
%		discussed the results and commented on the manuscript.
%%\item[Competing financial interests] The authors declare no
%%	competing financial interests.
%\item[Additional information] Correspondence and requests for materials should be
%	addressed to S.-H. Baek~(email: sbaek.fu@gmail.com).
%\end{description}
%
%\pagebreak

\end{document}